\documentclass[journal]{IEEEtran}
\ifCLASSINFOpdf
  \usepackage[pdftex]{graphicx}
  \graphicspath{{../plots/}{../figures/}}
  \DeclareGraphicsExtensions{.pdf,.jpeg,.png}
\else
  \usepackage[dvips]{graphicx}
  \graphicspath{{../plots/}{../figures/}}
  \DeclareGraphicsExtensions{.eps}
\fi
\usepackage[cmex10]{amsmath}
\usepackage{amssymb}

\newcommand{\BALA}[2]{\phi_2({\bf {#1}};#2)}
\newcommand{\PRemPHI}{\phi_2({\bf y} ;\tilde{\pi})}

\newcommand{\PPHI}[1]{\phi_2({\bf y} ;\tilde{\pi})}
\newcommand{\PPHII}[1]{\phi_2({\bf x}^{({#1})} ;\tilde{\pi})}
\newcommand{\SPHI}[1]{\phi_2({\bf X}^{({#1})} ;p)}

\newtheorem{theorem}{Theorem}
\newtheorem{lemma}{Lemma}

\newcommand{\sleep}{{${\sf sleep }$}}
\newcommand{\wake}{{${\sf awake }$}}

\newcommand{\pfa}{\mathsf{P_{FA}}}

\newcommand{\wsn}{$\mathsf{WSN}$}

\newcommand{\PROB}[1]{\mathsf{P}\left\{#1\right\}}

\newcommand{\EXP}[1]{\mathsf{E}\!\left[#1\right]}
\newcommand{\EXPm}[1]{\mathsf{E}_{\PRemPHI}\!\left[#1\right]}

\newcommand{\nn}{\nonumber \\}

\begin{document}
%
\title{Optimum Sleep--Wake Scheduling of Sensors\\ 
       for Quickest Event Detection	in\\
	   Small Extent Wireless Sensor Networks}
%
%
%

\author{K.~Premkumar$^\dagger$ 
    and~Anurag~Kumar$^\ddagger$
    \thanks{$\dagger$ K.~Premkumar's work on this paper was done during
	his doctoral work at the Indian Institute of Science, Bangalore, 
	India. He is currently with the Hamilton Institute, National
	University of Ireland, Maynooth, Ireland.
	E-mail: kprem@ece.iisc.ernet.in
	
	$\ddagger$ Anurag Kumar is with the Department of Electrical
	Communication Engineering, Indian Institute of Science, Bangalore --
	560 012, India. 
	E-mail: anurag@ece.iisc.ernet.in }
    \thanks{This is an expanded version of a paper that was presented in IEEE
	INFOCOM 2008. This work was supported by a project from DRDO, Government of India.}
    }

\maketitle

\begin{abstract}
We consider the problem of quickest event detection with \sleep--${\sf
wake}$ scheduling in small extent wireless sensor networks in which, at
each time slot, each sensor node in the ${\sf awake}$ state observes a
sample and communicates the information to the fusion centre. The sensor
nodes in the {\sleep} state do not sample or communicate any information
to the fusion centre, thereby conserving energy. At each time slot, the
fusion centre, after having received the samples from the sensor nodes
in the {\wake} state, makes a decision to ${\sf stop}$ (and thus declare
that the event has occurred) or to ${\sf continue}$ observing. If it
decides to ${\sf continue}$, the fusion centre also makes the decision
of choosing the number of sensor nodes to be in the {\wake} state in the
next time slot. We consider three alternative approaches to the problem
of choosing the number of sensor nodes to be in the {\wake} state in
time slot $k+1$, based on the information available at time slot $k$,
namely,
\begin{enumerate}
\item optimal control of $M_{k+1}$, the number of sensor nodes to be in 
      the {\sf awake} state in time slot $k+1$,
\item optimal control of $q_{k+1}$, the probability of a sensor node to 
	  be in the {\sf awake} state in time slot $k+1$, and 
\item optimal probability $q$ that a sensor node is in the {\sf awake} 
      state in any time slot.
\end{enumerate}
In each case, we formulate the problem as a sequential decision process.
We show that a sufficient statistic for detecting the event and choosing
an optimal control at time $k$ is the a posteriori probability of change
$\Pi_k$. Also, we show that the optimal stopping rule is a threshold
rule on the a posteriori probability of change. We provide a partial
characterisation of the optimal policies for choosing $M_{k+1}$ or
$q_{k+1}$, and then explore these policies numerically. The optimal
policy for $M_{k+1}$ can keep very few sensors {\wake} during the prechange
phase and then quickly increase the number of sensors in the {\wake}
state when a change is ``suspected.'' Among the three \sleep--${\sf
wake}$ algorithms described, we observe that the total cost is minimum
for the optimum control of $M_{k+1}$ and is maximum for the optimum
control on $q$.
\end{abstract}


\IEEEpeerreviewmaketitle

\section{Introduction}
\label{sec:introduction}
Event detection (e.g., physical intrusion of a human into a secure
region) is an important application of wireless sensor networks
({\wsn}s). Events for which such a {\wsn} is deployed are typically rare
events, and hence, much of the energy of the sensor nodes gets drained
away in the pre--event period. As sensor nodes are energy--limited
devices, this reduces the utility of the sensor network.  Thus, {\em in
addition to the problem of quickest event detection, we are also faced
with the problem of increasing the lifetime of sensor nodes} which we
address in this paper by means of optimal \sleep--${\sf wake}$
scheduling of sensor nodes.

A sensor node can be in one of two states, the {\sleep} state or the
{\wake} state. A node in the {\sleep} state conserves energy by
switching to a low--power state. In the {\wake} state, a sensor node can
make measurements, perform some computations, and then communicate
information to the fusion centre. For enhancing the utility and the
lifetime of the network, it is essential to have {\em optimal
\sleep--${\sf wake}$ scheduling} for the sensor nodes, while achieving
the measurement and the inference objective of the \wsn.

We are interested in the quickest detection of an event with a minimal
number of sensors in the {\wake} state. A common approach to this
problem is by having a fixed deterministic duty cycle for the
\sleep--${\sf wake}$ activity. However, the duty cycle approach does not
make use of the prior information about the event, nor the observations
made by the sensors, and hence is not optimal. 

Hence, in this paper, we formulate the problem as one of optimum
sequential change detection.  In the classical change detection problem
\cite{shiryayev63}, the decision maker after having observed each
sample, has to make a decision to ${\sf stop}$, or to ${\sf continue}$
observing the next sample. In such a situation, the decision maker is
concerned only about minimising the detection delay while keeping the
probability of false alarm bounded from above by $\alpha$, a parameter
of interest. However, in the kind of ${\sf WSN}$ application described
above, there is an additional cost associated with generating an
observation and communicating it to the decision maker, which we
incorporate in our formulation. To the best of our knowledge, our work
is the first to look at the problem of joint design of optimal change
detection and \sleep--${\sf wake}$ scheduling.

\subsection{Summary of Contributions}
\label{subsec:summary_of_contributions}
We summarise the main contributions of this paper below.
\begin{enumerate}
\item We provide a model for the \sleep--${\sf wake}$ scheduling of 
      sensors by taking into account the cost per observation (which is
	  the ${\sf sensing+computation+communication}$ cost) per sensor in
	  the {\wake} state and formulate the joint \sleep--${\sf wake}$
	  scheduling and quickest event detection problem subject to a false
	  alarm constraint, in the Bayesian framework, as an optimal control
	  problem. We show that the problem can be modelled as a partially 
	  observable Markov decision process (POMDP).

\item We obtain an average delay optimum stopping rule for event 
      detection and show that the stopping rule is a threshold rule on
	  the a posteriori probability of change. 

\item Also, at each time slot $k$, we obtain the optimal strategy for 
      choosing the optimum number of sensors to be in the {\wake} state in 
	  time slot $k+1$ based on the sensor observations until time $k$,
	  for each of the control strategies described as follows: 
      \begin{enumerate}
      \item[(i)]   control of $M_{k+1}$, the number of sensors to be in 
	               the {\wake} state in time slot $k+1$, 
      \item[(ii)]  control of $q_{k+1}$, the probability of a sensor to 
	               be in the {\wake} state in slot $k+1$, and
      \item[(iii)] constant probability $q$ of a sensor in the {\wake}
	               state in any time slot.
      \end{enumerate}
\end{enumerate}

\subsection{Discussion of the Related Literature}
\label{subsec:discussion_of_the_related_literature}
In this section, we discuss the most relevant literature on
energy--efficient detection. Censoring was proposed by Rago {\em et al.}
in \cite{rago-etal96censoring} as a means to achieve energy--efficiency.
{\em Binary hypothesis testing} with energy constraints was formulated
by Appadwedula {\em et al.} in
\cite{stat-sig-proc.appadwedula-etal05energy-efficient-detection}. These
schemes find the ``information content'' in any observation, and
uninformative observations are not sent to the fusion centre. Thus,
censoring saves only the communication cost of an observation. In our
work, by making a sensor go to the {\sleep} state, we save the ${\sf
sensing+computation+communication}$ cost of making an observation. 

In related work \cite{wu-etal07sleep-awake}, Wu {\em et al.} proposed a
low duty cycle strategy for \sleep--${\sf wake}$ scheduling for sensor
networks employed for data monitoring (data collection) applications. In
the case of sequential event detection, duty cycle strategies are not
optimal, and it would be beneficial to adaptively turn the sensor nodes
to the {\sleep} or {\wake} state based on the prior information, and the
observations made during the decision process, which is the focus of
this paper.

In \cite{lz-rs07decentralized}, Zacharias and Sundaresan studied the
problem of event detection in a ${\sf WSN}$ with physical layer fusion
and power control at the sensors for energy--efficiency. Their Markov
decision process (MDP) framework is similar to ours. However, in
\cite{lz-rs07decentralized}, all the sensor nodes are in the {\wake}
state at all time. In our work, we seek an optimal state dependent
policy for determining how many sensors to be kept in the {\wake} state,
while achieving the inference objectives (detection delay and false 
alarm).

\subsection{Outline of the paper}
\label{subsec:outline_of_the_paper}
The rest of the paper is organised as follows. In
Section~\ref{sec:problem_formulation}, we formulate the \sleep--${\sf
wake}$ scheduling problem for quickest event detection. We describe
various costs associated with the event detection problem. Also, we
outline various control strategies for \sleep--${\sf wake}$ scheduling
of sensor nodes. In
Section~\ref{sec:quickest_change_detection_with_feedback}, we discuss
the optimal \sleep--${\sf wake}$ scheduling problem that minimises the
detection delay when there is a feedback from the decision maker (in
this case, the fusion centre) to the sensors. In particular, the
feedback could be the number of sensors to be in the {\wake} state or
the probability of a sensor to be in the {\wake} state in the next time
slot. We show that the a posteriori probability of change is sufficient
for stopping and for controlling the number of sensors to be in the
{\wake} state. In
Section~\ref{sec:quickest_change_detection_without_feedback}, we discuss
an optimal open loop \sleep--${\sf wake}$ scheduler that minimises the
detection delay where there is no feedback from the fusion centre and
the sensor nodes. We obtain the optimal probability with which a sensor
node is in the {\wake} state at any time slot.  In
Section~\ref{sec:numerical_results}, we provide numerical results for
the \sleep--${\sf wake}$ scheduling algorithms we obtain.
Section~\ref{sec:conclusion} summarises the results in this paper.

\section{Problem Formulation}
\label{sec:problem_formulation}
In this section, we describe the problem of {\em quickest event
detection with a cost for taking observations} and set up the model. We
consider a {\sf WSN} comprising $n$ unimodal sensors (i.e., all the
sensors have the same sensing modality, e.g., acoustic, vibration,
passive infrared (PIR), or magnetic) deployed in a region $\mathcal{A}$
for an intrusion detection application. We consider a small extent 
network, i.e., the region $\mathcal{A}$ is covered by the {\em sensing 
coverage} of each of the sensors. An event (for example, a human
``intruder'' entering a secure space) happens at a random time. The
problem is to detect the event as early as possible with an optimal
\sleep--${\sf wake}$ scheduling of sensors subject to a false alarm
constraint.  

We consider a discrete time system and the basic unit of time is one
slot. The slots are indexed by non--negative integers. A time slot is
assumed to be of unit length, and hence, slot $k$ refers to the time
interval $[k,k+1)$. We assume that the sensor network is time
synchronised (see, \cite{solis-etal06time-synch} for achieving time
synchrony). An event occurs at a random time $T \in \mathbb{Z}_+$ and
persists from there on for all $k \geqslant T$. The prior distribution
of $T$ (the time slot at which the event happens) is given by
\begin{align*}
\PROB{T=k} & = \left\{
                    \begin{array}{ll}
                    \rho, & \ \text{ if } k = 0\\ 
    (1-\rho)(1-p)^{k-1}p, & \ \text{ if } k > 0,  
                    \end{array}
             \right.
\end{align*}
where $0 < p \le 1$ and $0 \le \rho \le 1$ represents the probability 
that the event happened even before the observations are made. We say 
that the state of nature, $\Theta_k$ is 0 before the occurrence of the 
event (i.e., $\Theta_k = 0$ for $k < T$) and 1 after the occurrence of 
the event (i.e., $\Theta_k = 1$ for $k \ge T$). 

At any time $k \in {\mathbb Z}_+$, the state of nature $\Theta_k$ can
not be observed directly and can be observed only partially through the
sensor observations. The observations are obtained sequentially starting
from time slot $k = 1$ onwards. Before the event takes place, i.e., for
$1 \le k < T$, sensor~$i$ observes $X_k^{(i)} \in \mathbb{R}$ the
distribution of which is given by $F_0(\cdot)$, and after the event
takes place, i.e., for $k \ge T$, sensor $i$ observes $X_k^{(i)} \in
\mathbb{R}$ the distribution of which is given by $F_1(\cdot) \ne
F_0(\cdot)$ (because of the small extent network, at time $T$, the
observations of all the sensors switch their distribution to the
postchange distribution $F_1(\cdot)$). The corresponding probability
density functions (pdfs) are given by $f_0(\cdot)$ and $f_1(\cdot) \neq
f_0(\cdot)$\footnote{If the observations are quantised, one can work
with probability mass functions instead of pdfs.}. Conditioned on the
state of nature, i.e., given the change point $T$, the observations
$X_k^{(i)}$s are independent across sensor nodes and across time. The
event and the observation models are essentially the same as in the
classical change detection problem, \cite{shiryayev} and
\cite{veeravalli01decentralized-quickest}. 

The observations are transmitted to a fusion centre. It is assumed that
the communication between the sensors and the fusion centre is
error--free and completes before the next measurements are
taken\footnote{This could be achieved by synchronous time division
multiple access, with robust modulation and coding. For a formulation
that incorporates a random access network (but not \sleep--${\sf wake}$
scheduling), see \cite{secon} and
\cite{wsn.premkumar-etal10det-over-mac}.}. At time $k$, let ${\cal M}_k
= \{i_{k,1}, i_{k,2}, \cdots, i_{k,M_k}\} \subseteq \{1,2,\cdots,n\}$ be
the set of sensor nodes that are in the {\wake} state, and the fusion
centre receives a vector of $M_k$ observations ${\bf Y}_k = {\bf
X}^{{\cal M}_k}_k := \big[X_k^{(i_{k,1})}, X_k^{(i_{k,2})}, \cdots,
X_k^{(i_{k,M_k})}\big]$. At time slot $k$, based on the observations so
far ${\bf Y}_{[1:k]}$,\footnote{The notation $Y_{[k_1:k_2]}$ 
defined for $k_1 \leq k_2$ means the
vector $[Y_{k_1}, Y_{k_1+1}, \cdots, Y_{k_2}]$. } the distribution of $T$, $f_0(\cdot)$, and
$f_1(\cdot)$, the fusion centre  
\begin{enumerate}
\item makes a decision on whether to raise an alarm or to continue 
      sampling, and 
\item if it decides to continue sampling, it determines the number of 
      sensors that must be in the {\wake} state in time slot $k+1$.
\end{enumerate}
Let $D_k \in \{0, 1\}$ be the decision made by the fusion centre to
``${\sf continue}$ ${\sf sampling}$'' in time slot $k+1$ (denoted by
0) or ``${\sf stop}$ ${\sf and}$ ${\sf raise}$ ${\sf an}$ ${\sf
alarm}$'' (denoted by 1). If $D_k = 0$, the fusion centre controls the
set of sensors to be in the {\wake} state in time slot $k+1$, and if
$D_k = 1$, the fusion centre chooses ${\cal M}_{k+1} = \emptyset$.  Let
$A_k \in \mathcal{A}$ be the decision (or control or action) made by the
fusion centre after having observed ${\bf Y}_k$ at time $k$. We note
that $A_k$ also includes the decision $D_k$. Also, the action space
$\mathcal{A}$ depends on the feedback strategy between the fusion centre
and the sensor nodes which we discuss in detail in
Section~\ref{sec:quickest_change_detection_with_feedback}. Let ${\bf
I}_k := [{\bf Y}_{[1:k]}, A_{[0,k-1]}]$ be the information available to
the decision maker at the beginning of slot $k$. The action or control
$A_k$ chosen at time $k$ depends on the information ${\bf I}_k$ (i.e.,
$A_k$ is ${\bf I}_k$ measurable).

The costs involved are i) $\lambda_s$, the cost due to
$\mathsf{(sampling + computation + communication)}$ per observation per
sensor, ii) $\lambda_f$, the cost of false alarm, and iii) the detection
delay, defined as the delay between the occurrence of the event and the
detection, i.e., $(\tau-T)^+$, where $\tau$ is the time instant at which
the decision maker {\sf stop}s sampling and raises an alarm\footnote{We
note here that the event $\{\tau=k\}$ is completely determined by the
information ${\bf I}_k$, and hence, $\tau$ is a stopping time with
respect to the sequence of random variables ${\bf I}_1, {\bf I}_2,
\cdots$.}. Let $c_k:\{0,1\}\times\{(0,0), (0,1), \cdots, (0,n),
(1,0)\}\to {\mathbb R}_+$ be the cost incurred at time slot $k$.  For $k
\leqslant \tau$, the one step cost function is defined (when the state
of nature is $\Theta_k$, the decision made is $D_k$, and the number of
sensors in the {\wake} state in the next time slot is $M_{k+1}$) as 
\begin{eqnarray}
& & c_k(\Theta_k,D_k,M_{k+1}) \nonumber \\
& := & 
    \left\{
    \begin{array}{ll}
	\lambda_s M_{k+1},       & \text{if} \ \Theta_k=0, D_k=0\\
	\lambda_f,         & \text{if} \ \Theta_k=0, D_k=1\\
	1+\lambda_s M_{k+1},     & \text{if} \ \Theta_k=1, D_k=0\\
	0,                 & \text{if} \ \Theta_k=1, D_k=1
    \end{array} 
	\right.
\end{eqnarray}
and for $k > \tau$, $c_k(\cdot,\cdot,\cdot) := 0$. Note that in the
above definition of the cost function, if the decision $D_k$ is 1, then
$M_{k+1}$ is always 0. For time $k \leqslant \tau$, the cost
$c_k(\Theta_k,D_k,M_{k+1})$ can be written as
\begin{eqnarray}
\label{eqn:general_cost_fn}
c_k(\Theta_k,D_k,M_{k+1}) 
\hspace{-3mm} & = & \hspace{-2mm} \lambda_f \cdot {\bf 1}_{\{\Theta_k = 0\}}  {\bf 1}_{\{D_k = 1\}} \nonumber \\
& & \hspace{-3mm}+ \left( {\bf 1}_{\{\Theta_k = 1\}} + \lambda_s M_{k+1} \right)  {\bf 1}_{\{D_k = 0\}}.
\end{eqnarray}

We are interested in obtaining a quickest detection procedure that
minimises the mean detection delay and the cost of observations by
sensor nodes in the {\wake} state subject to the constraint that the
probability of false alarm is bounded by $\alpha$, a desired quantity.
We thus have a constrained optimization problem,
\begin{eqnarray}
\label{eqn:constrained-opt-problem}
& & \text{minimise} \hspace{10mm} \EXP{(\tau-T)^+ + \lambda_s
\sum_{k=1}^\tau M_k} \ \ \ \ \ \\
& & \text{subject to} \hspace{8mm} \PROB{\tau < T} \le \alpha \nonumber
\end{eqnarray}
where $\tau$ is a stopping time with respect to the sequence ${\bf I}_1,
{\bf I}_2, \cdots$. The above problem would also arise if we imposed a
total energy constraint on the sensors until the stopping time (in which
case, $\lambda_s$ can be thought of as the {\em Lagrange multiplier} that
relaxes the energy constraint). 
Let $\lambda_f$ be the cost of false alarm. 
The expected total cost (or the Bayes risk) when the
stopping time is $\tau$ is given by 
\begin{eqnarray}
\label{eqn:unconstrained-problem}
R(\tau)  
& = & \lambda_f  \PROB{\tau<T} + \EXP{(\tau-T)^+  + \lambda_s  \sum_{k=1}^\tau M_k }\\
& = & \EXP{\lambda_f  {\bf 1}_{\left\{\Theta_\tau=0\right\}} + \sum_{k=0}^{\tau-1} \left( {\bf 1}_{\left\{\Theta_k=1\right\}} + \lambda_s  M_{k+1} \right)}\nonumber\\
& = & \EXP{c_\tau(\Theta_\tau, 1,0)  + \sum_{k=0}^{\tau-1} c_k(\Theta_k,0,M_{k+1}) }\nonumber\\
& = & \EXP{\sum_{k=0}^{\tau} c_k(\Theta_k,D_k,M_{k+1}) } \nonumber\\
& \stackrel{(a)}{=} & \EXP{\sum_{k=0}^{\infty} c_k(\Theta_k,D_k,M_{k+1})
}\nn
& \stackrel{(b)}{=} & \sum_{k=0}^{\infty} \EXP{c_k(\Theta_k,D_k,M_{k+1}) } 
\end{eqnarray}
where step $(a)$ follows from $c_k(\cdot,\cdot,\cdot)=0$ for $k >
\tau$, and step $(b)$ follows from the monotone convergence theorem. 
Note that $\lambda_f$ is a Lagrange multiplier and is chosen such that the false 
alarm constraint is satisfied with equality, i.e., $\pfa = \alpha$ 
(see \cite{shiryayev}). 

We note that the stopping
time $\tau$ is related to the control sequence $\{A_k\}$ in the
following manner. For any stopping time $\tau$, there exists a sequence
of functions (also called a policy) $\nu=(\nu_1,\nu_2,\cdots)$ such that
for any $k$, when $\tau=k$, $D_{k'} = \nu_{k'}({\bf I}_{k'}) = 0$ for 
all $k' < k$ and $D_{k'} = \nu_{k'}({\bf I}_{k'}) = 1$ for all $k' 
\geqslant k$. Thus, the unconstrained expected cost given by Eqn.~\ref{eqn:unconstrained-problem}  
is 
\begin{eqnarray}
 R(\tau)
& = & \sum_{k=0}^{\infty} \EXP{c_k(\Theta_k,D_k,M_{k+1}) } \nn
& = & \sum_{k=0}^{\infty} \EXP{c_k(\Theta_k,\nu_k({\bf I}_k),M_{k+1}) } \nn
& = & \sum_{k=0}^{\infty} \EXP{\EXP{c_k(\Theta_k,\nu_k({\bf I}_k),M_{k+1})\mid {\bf
I}_k }}\nn
 \label{opt-prob}
& \stackrel{(a)}{=} & \EXP{\sum_{k=0}^{\infty} \EXP{c_k(\Theta_k,\nu_k({\bf I}_k),M_{k+1})\mid {\bf
I}_k }}  \\
& = & \EXP{\sum_{k=0}^{\tau} \EXP{c_k(\Theta_k,\nu_k({\bf I}_k),M_{k+1})\mid {\bf I}_k
}}\nonumber
\end{eqnarray}
where step $(a)$ above follows from the monotone convergence theorem. 
From Eqn.~\ref{eqn:general_cost_fn}, it is clear that for $k \leqslant \tau$
\begin{eqnarray*}
& & \EXP{c_k(\Theta_k,\nu_k({\bf I}_k),M_{k+1})\mid {\bf I}_k } \nn
& = & \EXP{\lambda_f \cdot {\bf 1}_{\{\Theta_k = 0\}}
\cdot {\bf 1}_{\{\nu_k({\bf I}_k) = 1\}}} \nn
& & + \EXP{\left( {\bf 1}_{\{\Theta_k = 1\}} + \lambda_s M_{k+1} \right) \cdot
{\bf 1}_{\{\nu_k({\bf I}_k) = 0\}}
\mid {\bf I}_k }\nn
& = & \lambda_f \cdot \EXP{{\bf 1}_{\{\Theta_k = 0\}} \mid {\bf I}_k } 
\cdot {\bf 1}_{\{\nu_k({\bf I}_k) = 1\}} \nn
& & + 
\left(
\EXP{ {\bf 1}_{\{\Theta_k = 1\}} \mid {\bf I}_k } + 
\lambda_s \cdot \EXP{ M_{k+1}  \mid {\bf I}_k } \right) \cdot {\bf
1}_{\{\nu_k({\bf I}_k) = 0\}}
\end{eqnarray*}
For $k \leqslant \tau$, define the a posteriori probability of the change having occurred 
at or before time slot $k$, 
$\Pi_k := \EXP{{\bf 1}_{\{\Theta_k = 1\}} \big\arrowvert {\bf I}_k}$, and hence, we
have 
\begin{eqnarray}
\label{eqn:ctilde-cost-fn}
&  & \EXP{c_k(\Theta_k,\nu_k({\bf I}_k),M_{k+1})\mid {\bf I}_k } \nn
& = & \lambda_f \cdot (1-\Pi_k)  {\bf 1}_{\{\nu_k({\bf I}_k) = 1\}}\nn
& & + 
\left( \Pi_k + \lambda_s \cdot \EXP{ M_{k+1}  \mid {\bf I}_k } \right)
 {\bf 1}_{\{\nu_k({\bf I}_k) = 0\}}.
\end{eqnarray}
Thus, we can write the Bayesian risk given in
Eqn.~\ref{opt-prob} as 
\begin{eqnarray}
R(\tau) = \EXP{\lambda_f \cdot (1-\Pi_\tau) + \sum_{k=0}^{\tau-1}
\left( \Pi_k + \lambda_s \EXP{ M_{k+1} \mid {\bf I}_k} \right)} 
\end{eqnarray}
We are interested in obtaining an optimal stopping time $\tau$ and an
optimal control of the number of sensors in the {\wake} state. Thus, we
have the following problem, 
\begin{eqnarray}
\label{eqn:bridge}
\text{minimise} \ \EXP{
\lambda_f \cdot (1-\Pi_\tau) + \sum_{k=0}^{\tau-1}
\left( \Pi_k + \lambda_s \EXP{ M_{k+1} \mid {\bf I}_k} \right)} 
\hspace{-2mm}
\end{eqnarray}
We consider the following possibilities for the problem defined in Eqn.~\ref{eqn:bridge}.
\begin{enumerate}
\item {\bf Closed loop control on $M_{k+1}$}: 
      At time slot $k$, the fusion centre makes a decision on $M_{k+1}$, the number of 
      sensors in the {\wake} state in time slot $k+1$, based on the information 
      available (at the fusion centre) up to time slot $k$. The decision is then 
      fed back to the sensors via a feedback channel. Thus, the problem becomes
      \begin{eqnarray}
      \label{eqn:closed-cost-fn-M}
	    \min_{\tau, M_1, M_2, \cdots, M_\tau} 
       \EXP{ \lambda_f  (1-\Pi_{\tau}) + \sum_{k=0}^{\tau-1} \left(\Pi_k
	   + \lambda_s  M_{k+1}\right) } \hspace{-2mm}
      \end{eqnarray}

\item {\bf Closed loop control on $q_{k+1}$}: 
      At time slot $k$, the fusion centre makes a decision on $q_{k+1}$, the probability
      that a sensor is in the {\wake} state at time slot $k+1$ based on
	  the information ${\bf I}_k$. $q_{k+1}$ is 
      then broadcast via a feedback channel to the sensors. Thus, given
	  ${\bf I}_k$,
      the number of sensors in the {\wake} state $M_{k+1}$, at time slot $k+1$, is
      Bernoulli distributed with parameters $(n,q_{k+1})$ and 
	  $\EXP{M_{k+1}\mid {\bf I}_k} = nq_{k+1}$. 
	  Thus, the problem defined in Eqn.~\ref{eqn:bridge} becomes  
      \begin{eqnarray}
      \label{eqn:closed-cost-fn-q}
        \min_{\tau, q_{1}, q_2 \cdots,q_\tau} 
      \EXP{\lambda_f  (1-\Pi_{\tau}) + 
                      \sum_{k=0}^{\tau-1} \left(\Pi_k + \lambda_s
					  nq_{k+1}\right)} \hspace{-2mm} 
      \end{eqnarray}

\item {\bf Open loop control on $q$}: Here, there is no feedback between
      fusion centre and the sensor nodes. 
      At time slot $k$, each sensor node is in the {\wake} state with 
	  probability $q$. Note that $M_k$, the number of sensors in the
	  {\wake} state at time slot $k$ is Bernoulli distributed with
	  parameters $(n,q)$. Also note that $\{M_k\}$ process is i.i.d.\
	  and $\EXP{M_{k+1}\mid {\bf I}_k} = nq$ (also, $M_{k+1}$
	  is independent of the information vector ${\bf I}_k$). Note that
	  {\em the probability $q$ is constant over time}. Thus, the problem
	  defined in Eqn.~\ref{eqn:bridge} becomes 
      \begin{eqnarray}
      \label{eqn:open-cost-fn}
       \min_{\tau}  \EXP{ \lambda_f  (1-\Pi_{\tau}) + \sum_{k=0}^{\tau-1} \left(\Pi_k
	   + \lambda_s  nq\right)}
      \end{eqnarray}
      Here, $q$ is chosen (at time $k=0$) such that it minimises the above cost.  
\end{enumerate}
Note that the first two scenarios require a feedback channel between the
fusion centre and the sensors whereas the last scenario does not require
a feedback channel.

In Section~\ref{sec:quickest_change_detection_with_feedback}, we
formulate the optimization problem defined in
Eqns.~\ref{eqn:closed-cost-fn-M} and \ref{eqn:closed-cost-fn-q} in
the framework of MDP and study the optimal
closed loop \sleep--${\sf wake}$ scheduling policies. In
Section~\ref{sec:quickest_change_detection_without_feedback}, we
formulate the optimization problem defined in
Eqn.~\ref{eqn:open-cost-fn} in the MDP framework and obtain the
optimal probability $q$ of a sensor in the {\wake} state. 

\section{Quickest Change Detection with Feedback}
\label{sec:quickest_change_detection_with_feedback}
In this section, we study the \sleep--${\sf wake}$ scheduling problem when 
there is feedback from the fusion centre to the sensors. 

At time slot $k$, the fusion centre receives a $M_k$--vector of
observations ${\bf Y}_k$, and computes $\Pi_k$. Recall that $\Pi_k =
\mathsf{P}\left\{T\le k \Big \arrowvert {\bf I}_k\right\}$ is the a
posteriori probability of the event having occurred at or before time
slot $k$. For the event detection problem, a sufficient statistic for
the sensor observations at time slot $k$ is given by $\Pi_k$ (see
\cite{ss} and page 244, \cite{books.bertsekas00a}). When an {\em alarm} is raised, the system enters into a
terminal state `$\mathsf{t}$'. Thus, the state space of the $\{\Pi_k\}$
process is  ${\mathcal S} = [0, 1] \cup \{ \mathsf{t} \}$.
Note that $\Pi_k$ is also called the {\em information state} of the system. 

In the rest of the section, we explain the MDP formulation that yields
the closed loop \sleep--${\sf wake}$ scheduling algorithms. 

\subsection{Control on the number of sensors in the {\wake} state}
In this subsection, we are interested in obtaining an optimal control on 
$M_{k+1}$, the number of sensors in the {\wake} state, based on the
information we have at time slot $k$. 

At time slot $k$, after having
observed ${\bf X}^{{\cal M}_k}_k$, the fusion centre computes the
sufficient statistic $\Pi_k$. Based on $\Pi_k$, the fusion centre makes
a decision to ${\sf stop}$ or to ${\sf continue}$ sampling. If the decision
is to ${\sf continue}$ at time slot $k+1$, the fusion centre (which also
acts as a controller) chooses $M_{k+1}$, the number of sensors to be
in the {\wake} state at time slot $k+1$. The fusion centre also keeps
track of the residual energy in the sensor nodes, based on which it 
chooses the set of sensor nodes ${\cal M}_{k+1}$ that must be in the 
{\wake} state in time slot $k+1$. Since, the prechange and the 
postchange pdfs of the observations are the same for all the 
sensor nodes and at any time, the sensor observations are conditionally independent 
across sensors, any observation vector of size $m$ has the same pdf and 
hence, for decision making, it is sufficient to look at only the number 
of sensors in the {\wake} state $M_{k+1}$, i.e., the indices of the sensor nodes 
that are in the {\wake} state are not required for detection 
(we assume that the fusion centre chooses the sequence ${\cal M}_1, {\cal M}_2, \cdots$ in such a
way that the rate at which the sensor nodes drain their energy is the
same). Thus, the set of controls at time slot $k$ is given by 
${\mathcal A} = \bigg\{ (\mathsf{stop},0),
\bigcup_{m\in\{0,1,\cdots,n\}}(\mathsf{continue},m) \bigg\}$
$=\big\{(1,0), (0,0), (0,1), \cdots, (0,n)\big\}$.

We show that $\Pi_k$ can be computed in a recursive manner from the 
previous state $\Pi_{k-1}$, the previous action $A_{k-1}$, and the 
current observation ${\bf X}^{{\cal M}_k}_k$ as, 
\begin{eqnarray}
\label{eqn:recursion} 
& & \Pi_{k} \nn
&  = & {\Phi}(\Pi_{k-1},A_{k-1}, {\bf X}^{{\cal M}_k}_k)\nn 
         & := & \left\{
			 \begin{array}{ll}
             {\sf t}, & \text{if} \ \Pi_{k-1} = {\sf t}\\
             {\sf t}, & \text{if} \ A_{k-1} = 1 \\
             \frac{\widetilde{\Pi}_{k-1}\phi_1\left({\bf X}^{{\cal
			 M}_k}_k\right)
}{\phi_2\left({\bf X}^{{\cal M}_k}_k; 	\widetilde{\Pi}_{k-1}\right) }, &
\text{if} \ \Pi_{k-1} \in [0, 1],  A_{k-1} = (0,M_k) 
			 \end{array}
			 \right.
\end{eqnarray} 
where  
\begin{eqnarray}
\label{eqn:some-definitions}
\widetilde{\Pi}_k & := & \Pi_k+(1-\Pi_k)p, \nn 
\phi_0\left({\bf X}_k^{{\cal M}_k}\right)  & := & \prod_{i \in {\cal M}_k} f_0(X_k^{(i)}), \nonumber \\ 
\phi_1\left({\bf X}_k^{{\cal M}_k}\right)  & := & \prod_{i \in {\cal M}_k} f_1(X_k^{(i)}), \nonumber \\ 
\phi_2\left({\bf X}_k^{{\cal M}_k};\widetilde{\Pi}\right)  & := & \widetilde{\Pi}
\phi_1\left({\bf X}_k^{{\cal M}_k}\right) + 
(1-\widetilde{\Pi})
\phi_0\left({\bf X}_k^{{\cal M}_k}\right) \ \ \ \ \  
\end{eqnarray}
Thus, the a posteriori probability process $\{\Pi_k\}$ is a 
controlled Markov process. Note that $\widetilde{\Pi}_k = \Pi_k + (1-\Pi_k)p =  
\EXP{\Pi_{k+1}}$ before ${\bf X}_{k+1}^{{\cal M}_{k+1}}$ is observed. 
Motivated by the structure of the cost given in 
Eqn.~\ref{eqn:ctilde-cost-fn}, we define the 
one stage cost function
$\widetilde{c}:\mathcal{S}\times\mathcal{A}\to\mathbb{R}_+$ 
when the (state, action) 
pair is $(s, a)$ 
as
\begin{eqnarray*}
\widetilde{c}(s,a) 
& = & \left\{
         \begin{array}{ll}
         \lambda_f \left(1-\pi\right), & 
		 \text{if} \ s=\pi \in [0,1], a = (1,0)\\ 
         \pi + \lambda_s m ,           & \text{if} \ s=\pi \in [0,1], a = (0,m)\\
         0,                            & \text{if} \ s = \mathsf{t}.
         \end{array}
      \right.
\end{eqnarray*}
Since $M_{k+1}$ is chosen based on the information ${\bf I}_k$, there 
exists a function $\nu_k'$ such
that $M_{k+1} = \nu_k'({\bf I}_k)$. Thus, the action or control at time
$k$ is given by $\mu_k({\bf I}_k) = \left(\nu_k({\bf I}_k), \nu_k'({\bf
I}_k)\right)$. Hence, we can write the Bayesian risk given in 
Eqn.~\ref{eqn:unconstrained-problem} for a policy
$\mu=(\mu_1,\mu_2,\cdots)$ as 
\begin{eqnarray}
R(\tau) & = & \EXP{\sum_{k=0}^\infty \widetilde{c}\left(\Pi_k,\mu_k({\bf I}_k)\right)}\nn
& = & \EXP{\sum_{k=0}^\infty \widetilde{c}\left(\Pi_k,\widetilde{\mu}_k({\Pi}_k)\right)} 
\end{eqnarray}
Since $\Pi_k$ is a sufficient statistic for ${\bf I}_k$, for any policy $\mu_k$
there exists a corresponding policy $\widetilde{\mu}_k$ such that 
$\widetilde{\mu}_k(\Pi_k) = {\mu}_k({\bf I}_k)$, and hence, the  
last step in the above equation follows (see page 244, \cite{books.bertsekas00a}) 
Since, the one stage cost and the density function 
$\phi_2({\bf y};\widetilde{\Pi}_{k-1})$ are time invariant, it is sufficient to consider the
class of stationary policies 
(see Proposition 2.2.2 of \cite{books.bertsekas07b}). Let $\widetilde\mu:\mathcal{S}\to\mathcal{A}$ be a 
stationary policy. Hence, the cost of using the policy $\widetilde\mu$ is given by 
\begin{eqnarray*}
J_{\widetilde\mu}(\pi_0) & = &  \EXP{\sum_{k=0}^\infty \widetilde c(\Pi_k,\widetilde\mu(\Pi_k)) \bigg\arrowvert \Pi_0 = \pi_0},
\end{eqnarray*}
and hence, the minimal cost among the class of stationary
policies is given by
\begin{eqnarray*}
J^*(\pi_0) & = & \min_{\widetilde\mu} \EXP{\sum_{k=0}^\infty  \widetilde c(\Pi_k,\widetilde\mu(\Pi_k))
\bigg\arrowvert \Pi_0 = \pi_0}. 
\end{eqnarray*}
The dynamic program (DP) that solves the above problem is given by the Bellman's equation,
\begin{eqnarray}
\label{eqn:opt_J*_M*} 
J^*(\pi) & = & \min\bigg\{\widetilde c(\pi,1), H_{J^*}(\pi) \bigg\}
\end{eqnarray}
where the function $H_{J^*}:[0,1]\to\mathbb{R}_+$ is defined as
\begin{eqnarray}
& & H_{J^*}(\pi)\nn 
& := & \hspace{-5mm}\min_{0\le m \le
n} \hspace{-3mm}\left\{ \widetilde c(\pi,(0,m)) + \EXPm{
J^*\left(
\Phi(\pi,(0,m),{\bf Y})
\right) 
} \right\} \ \ \ \ 
\end{eqnarray}
where ${\bf Y}$ and ${\bf y}$ are $m$--vectors.
The notation ${\mathsf E}_{\PRemPHI}[\cdot]$ means that 
the expectation is taken with respect to the pdf $\PRemPHI$
(recall Eqn.~\ref{eqn:some-definitions} for the definition of $\PRemPHI$). 
Thus, Eqn.~\ref{eqn:opt_J*_M*} can be written as 
\begin{eqnarray}
\label{eqn:opt_M_staionary_policy} 
J^*(\pi)  
& = & \min\left\{\lambda_f\cdot\big(1-\pi\big), \pi+  A_{J^*}\big(\pi\big)\right\}
\end{eqnarray}
where the function $A_{J^*}:[0,1]\to\mathbb{R}_+$ is defined as
\begin{eqnarray}
\label{eqn:OPT_CONTROL_M_A_J_STAR} 
A_{J^*}(\pi) = \min_{0\le m \le
n}\left\{\lambda_sm + \EXPm{
J^*\left(\frac{\tilde{\pi}\cdot\phi_1({\bf Y})}{\phi_2({\bf Y};\tilde{\pi})}\right) 
} \right\} 
\end{eqnarray}
The optimal policy $\mu^*$ that achieves $J^*$ gives the optimal
stopping rule, $\tau^*$, and the optimal number of sensors in the
{\wake} state, 
$M_1^*, 
M_2^*, \cdots, 
M_{\tau^*}^*$. 

\vspace{5mm}

\noindent
We now establish some properties of the {\em minimum} total cost
function $J^*$.

\begin{theorem}
\label{thm:opt_m_J*_concave}
The total cost function $J^*({\pi})$ is concave in $\pi$. 
\end{theorem}

\vspace{5mm}

\noindent
Also, we establish some properties of the 
optimal policy $\mu^*$ (which
maps the a posteriori probability of change $\Pi_k$ to the action
space ${\cal A}$) in the next theorem. 

\begin{theorem}
\label{thm:policy-M}
The optimal stopping rule is given by the following threshold rule
where the threshold is on the a posteriori probability of change,
\begin{eqnarray}
\tau^* & = & \inf\{k: \Pi_k \geqslant \Gamma\},
\end{eqnarray}
for some $\Gamma \in [0, 1]$. The threshold $\Gamma$ depends on the
probability of false alarm constraint, $\alpha$ (among other parameters
like the distribution of $T$, $f_0$, $f_1$). 
\end{theorem}

Theorem~\ref{thm:policy-M} addresses only the {\em stopping time} 
part of the optimal policy $\mu^*$. We now explore the structure of the 
optimal closed loop control policy for $M^*:[0, 1]\to\mathbb{Z}_+$, the 
optimal number of sensors in the {\wake} state in the {\em next} time 
slot. At time $k$, based on the (sufficient) statistic $\Pi_k$, the 
fusion centre chooses $M^*_{k+1}=M^*(\Pi_k)$ number of sensor nodes in
the {\wake} state. For each $0 \le m \le n$, we define the functions 
$B_{J^*}^{(m)}:[0,1]\to\mathbb{R}_+$ and 
$A_{J^*}^{(m)}:[0,1]\to\mathbb{R}_+$ as
\begin{eqnarray*} 
B_{J^*}^{(m)}(\pi) & := & {\mathsf
E}_{\PPHI{m}}\left[J^*\left(\frac{\tilde{\pi}\cdot\phi_1({\bf Y})}{\phi_2({\bf Y};\tilde{\pi})}\right)\right], \\ 
\text{and } A_{J^*}^{(m)}(\pi) & := & \lambda_sm + B_{J^*}^{(m)}({\pi}). 
\end{eqnarray*}
We have shown in the proof of Theorem~\ref{thm:opt_m_J*_concave}
that for any $m=0,1,2,\cdots,n$, the functions $B_{J^*}^{(m)}(\pi)$ 
and $A_{J^*}^{(m)}(\pi)$ are concave in $\pi$.

\begin{theorem}
\label{thm:monotone}
For any $\pi \in [0,1]$, the functions
$B_{J^*}^{(m)}(\pi)$ monotonically decrease with $m$. 
\end{theorem}

\noindent
{\bf Remark:} By increasing the number of sensor nodes in the 
{\wake} state, i.e., by
increasing $m$, we expect that the a posteriori probability of
change will get closer to 1 or closer to 0 (depending on whether the
change has occurred or not). In either case, the
one stage cost decreases, and hence, we expect that the functions   
$B_{J^*}^{(m)}(\pi)$ monotonically decrease with $m$.

\begin{figure}
\centering
\includegraphics[width=3.5in]{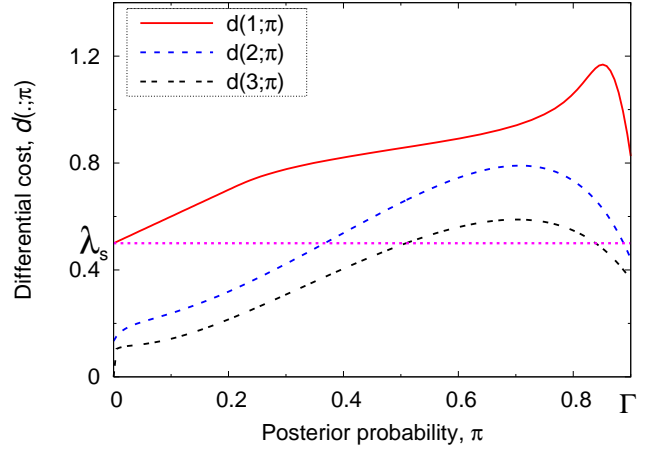}
\caption{Differential costs, $d(\cdot;\pi)$, for $n = 10$ sensors, $\lambda_f = 100.0$, $\lambda_s = 0.5$, $f_0 \sim \mathcal{N}(0,1)$ and $f_1 \sim \mathcal{N}(1,1)$.}
\label{fig:Opt_control_M_D}
\end{figure}

At time $k$, $B_{J^*}^{(m)}(\Pi_k)$ can be thought of as the cost--to--go 
function from slot $k+1$ onwards (having used $m$ sensor nodes at time
$k+1$). Note
that $A_{J^*}^{(m)}(\pi)$ has two components, the first component
$\lambda_s m$ increases with $m$ and (from 
Theorem~\ref{thm:monotone}) the second component decreases with $m$.
As $m$ takes values in a finite set $\{0,1,2,\cdots,n\}$, for each $\pi$, there exists an optimal $M^*(\pi)$ for which
$A_{J^*}^{(M^*(\pi))}(\pi)$ is minimum. 
For any given $\pi \in [0,1]$, we define the differential cost  
$d:\{1,2,\cdots,n\} \to \mathbb{R}_+$ as 
\begin{eqnarray}
\label{eqn:differential_cost_function}
d(m;\pi) & = & B_{J^*}^{(m-1)}(\pi) - B_{J^*}^{(m)}(\pi). 
\end{eqnarray}
Note that for any $1 \le m \le n$, $d(m;\pi)$ is bounded and continuous
in $\pi$ (as $B_{J^*}^{(m)}$s are bounded and concave in $\pi$). Also
note that $d(m;1) = 0$ as $B_{J^*}^{(m-1)}(1) = B_{J^*}^{(m)}(1) = 0$. We are interested
in  $d(m;\pi)$ for $\pi \in [0, \ \Gamma)$.
In Figure~\ref{fig:Opt_control_M_D}, we plot 
$d(m;\pi)$ against $\pi$ for $m=1,2,$ and 3 (for the set of parameters 
$n = 10$, $\lambda_f = 100$, $\lambda_s = 0.5$, and $f_0$ and $f_1$ are unit variance Gaussian 
pdfs with means 0 and 1 respectively). We observe that 
$d(m;\pi)$ monotonically decreases in $m$, for each $\pi \in [0,\Gamma)$
(i.e., $d(1;\pi) \ge d(2;\pi) \ge d(3;\pi)$).
We have observed this monotonicity property for different sets of experiments for the 
case when $f_0$ and $f_1$ belong to the Gaussian class of distributions. 
We conjecture that this monotonicity property of $d$ holds and state the following theorem
which gives a {\em structure} for $M^*$, the optimal number of sensors in the {\wake}
state.  
\begin{theorem}
\label{thm:label1}
If for each $\pi \in [0,\Gamma)$, $d(m;\pi)$ decreases monotonically in 
$m$, then the optimal number of sensors in the {\wake} state, 
$M^*:[0,1]\rightarrow \{0,1,\cdots,n\}$ is given by
\begin{eqnarray*}
M^*(\pi) & = & \max \big\{m: d(m;\pi) \ge \lambda_s\big\}.
\end{eqnarray*}
\end{theorem}

\subsection{Control on the probability of a sensor in the {\wake} state}
In this subsection, we are interested in obtaining an optimal control on 
$q_{k+1}$, the probability of a sensor in the {\wake} state, based on the
information we have at time slot $k$, instead of determining the number 
of sensors that must be in the {\wake} state in the next slot. 

At time slot $k$, after having observed ${\bf X}_k^{{\cal M}_k}$, the 
fusion centre computes the sufficient statistic $\Pi_k$, based on which
it makes a decision to ${\sf stop}$ or to ${\sf continue}$ sampling. If the 
decision is to ${\sf continue}$ at time slot $k+1$, the fusion centre 
(also acts as a controller) chooses $q_{k+1}$, the probability of a 
sensor to be in the {\wake} state at time slot $k+1$. Thus, the set of 
controls at time slot $k$ is given by 
$\mathcal{A}~=~\bigg\{
(\mathsf{stop},0), \cup_{q \in [0,1]} (\mathsf{continue},q) \bigg\}$ = $\bigg\{1, 
\cup_{q \in [0,1]} (0,q)\bigg\} = \{(1,0), \{0\}\times[0, 1]\}$.

When the control $A_k = (0,q_{k+1})$ is chosen, $M_{k+1}$, the number of sensors in the 
{\wake} state at time slot $k+1$ is {\em Bernoulli} distributed with parameters 
$(n,q_{k+1})$. Let $\gamma_m(q_{k+1})$ be the probability that $m$ sensors are in the 
{\wake} state at time slot $k+1$. $\gamma_m(q_{k+1})$ is given by 
\begin{eqnarray}
\gamma_m(q_{k+1}) & = & \binom{n}{m}q_{k+1}^m (1-q_{k+1})^{n-m}.
\end{eqnarray}
The information state at time slot $k$ $\Pi_k$, can be computed in a recursive
manner from $\Pi_{k-1}$, $A_{k-1}$ and ${\bf X}_k^{{\cal M}_k}$ 
using Eqn.~\ref{eqn:recursion}. Thus, it is clear that 
the $\{\Pi_k\}$ process is a controlled Markov process, the 
state space of the process being $\mathcal{S} = [0,1] \cup \{{\sf t}\}$.  
Motivated by the cost function given in 
Eqn.~\ref{eqn:ctilde-cost-fn}, define the one stage cost 
function $\widetilde c\left(s,a\right)$ 
when the (state,action) pair is $(s,a)$ as
\begin{eqnarray*}
\widetilde c\left(s,a\right)
& = &  \left\{
                           \begin{array}{ll}  
                           \lambda_f (1 - \pi), & \text{if} \ s=\pi \in [0,1], a =
						   (1,0)\\
                           \pi + \lambda_s nq,  & \text{if} \ s=\pi \in [0,1], a = (0,q)\\
                           0,                   & \text{if} \ s = \mathsf{t}.
                           \end{array} 
                           \right.    
\end{eqnarray*}
Since, the one stage cost and the density function 
$\phi_2({\bf y};\widetilde{\Pi}_{k-1})$ are time invariant, it is sufficient to consider the
class of stationary policies
(see Proposition 2.2.2 of \cite{books.bertsekas07b}). Let $\widetilde\mu:\mathcal{S}\to\mathcal{A}$ be a 
stationary policy. Hence, the cost of using the policy $\widetilde\mu$ is given by 
\begin{eqnarray*}
J_{\widetilde\mu}(\pi_0) & = &  \EXP{\sum_{k=0}^\infty \widetilde c(\Pi_k,\widetilde\mu(\Pi_k)) \bigg\arrowvert \Pi_0 = \pi_0},
\end{eqnarray*}
and hence the minimal cost among the class of stationary policies is
given by
\begin{eqnarray*}
J^*(\pi_0) & = & \min_{\widetilde\mu} \EXP{\sum_{k=0}^\infty \widetilde c(\Pi_k,\widetilde\mu(\Pi_k))
\bigg\arrowvert \Pi_0 = \pi_0}. 
\end{eqnarray*}
The DP that solves the above problem is given by the Bellman's equation,
\begin{eqnarray*}
J^*(\pi) & = & \min\left\{\widetilde c(\pi,1), H_{J^*}(\pi)\right\}
\end{eqnarray*}
where $H_{J^*}:[0,1] \to {\mathbb R}_+$ is defined as 
{\footnotesize
\begin{eqnarray*}
& & H_{J^*}(\pi) \\
& :=&   \min_{0 \le q \le 1}\left\{ \widetilde c(\pi,(0,q))+
\sum_{m=0}^n\gamma_m(q)\EXPm{J^*\big(\Phi(\pi,(0,m),{\bf Y}\big)}\right\}
\end{eqnarray*}
}
where ${\bf Y}$ and ${\bf y}$ are $m$--vectors.
Recall that the expectation is taken with respect to the pdf $\PRemPHI$.
The Bellman's equation can be written as 
\begin{eqnarray} 
\label{eqn:DP_for_control_q_k}
J^*(\pi)  
 & = & \min\left\{\lambda_f\cdot\big(1-\pi\big),  \pi + A_{J^*}\big(\pi\big)\right\}
\end{eqnarray}
where the function $A_{J^*}:[0,1]\to\mathbb{R}_+$ is defined as
\begin{eqnarray*} 
& & A_{J^*}(\pi) \\
& = & \min_{q \in [0,1]}\left\{\lambda_snq + \sum_{m=0}^n \gamma_m(q)
\EXPm{
J^*\left(\frac{\tilde{\pi}\cdot\phi_1({\bf Y})}{\phi_2({\bf Y};\tilde{\pi})}\right) 
} \right\}. 
\end{eqnarray*}
The optimal policy $\mu^*$ 
gives the optimal
stopping time $\tau^*$, 
and the optimal probabilities, $q_k^*, \ k~=~1,2,\cdots,\tau^*$.
The structure of the optimal policy is shown in the following theorems.

\begin{theorem}
\label{thm:label3}
The total cost function $J^*(\pi)$ is concave in $\pi$. 
\end{theorem}

\begin{theorem}
\label{thm:policy}
The optimal stopping rule is a threshold rule where the threshold is on
the a posteriori probability of change,
\begin{eqnarray*}
\tau^* & = & \inf\{k : \Pi_k \geqslant \Gamma\},
\end{eqnarray*}
for some $\Gamma \in [0, 1]$. The threshold $\Gamma$ depends on the
probability of false alarm constraint, $\alpha$ (among other parameters
like the distribution of $T$, $f_0$, $f_1$). 
\end{theorem}

\section{Quickest Change Detection without Feedback}
\label{sec:quickest_change_detection_without_feedback}
In this section, we study the \sleep--${\sf wake}$ scheduling problem
defined in Eqn.~\ref{eqn:open-cost-fn}. Open loop control is applicable
to the systems in which there is no feedback channel from the fusion
centre (controller) to the sensors. Here, at any time slot $k$, a sensor
chooses to be in the {\wake} state with probability $q$ independent of
other sensors. Hence, $\{M_k\}$, the number of sensors in the {\wake}
state at time slot $k$ is i.i.d. {\em Bernoulli distributed} with
parameters $(n,q)$. Let $\gamma_m$ be the probability that $m$ sensors
are in the {\wake} state. $\gamma_m$ is given by
\begin{eqnarray}
\gamma_m & = & \binom{n}{m}q^m (1-q)^{n-m}
\end{eqnarray}
We choose $q$ that minimises the Bayesian cost given by Eqn.~\ref{eqn:open-cost-fn}.

At time slot $k$, the fusion centre receives a vector of observation
${\bf X}_k^{{\cal M}_k}$ and computes $\Pi_k$. In the open loop
scenario, the state space is $\mathcal{S} =
\big\{[0,1]\cup\{\mathsf{t}\}\big\}$.  The set of actions is given by
$\mathcal{A} = \{\mathsf{stop},\mathsf{continue}\} =  \{1,0\}$  where
`1' represents ${\sf stop}$ and `0' represents ${\sf continue}$. Note that
$\Pi_k$ can be computed from $\Pi_{k-1}, A_{k-1}$, and ${\bf X}_k^{{\cal
M}_k}$ in the same way as shown in Eqn.~\ref{eqn:recursion}. Thus,
$\{\Pi_k\}$, $k \in \mathbb{Z}_+$ is a controlled Markov process.
Motivated by the structure of the cost given in
Eqn.~\ref{eqn:ctilde-cost-fn}, we define the 
one stage cost function
$\widetilde{c}:\mathcal{S}\times\mathcal{A}\to\mathbb{R}_+$ 
when the (state, action)
pair is $(s, a)$ as
\begin{eqnarray*}
\widetilde c(s,a) & = & \left\{
                  \begin{array}{ll}
                  \lambda_f(1-\pi),   & \text{if} \ s=\pi \in [0,1], a = 1\\
                  \pi + \lambda_s nq, & \text{if} \ s=\pi \in [0,1], a = 0\\
                  0,                  & \text{if} \ s = \mathsf{t}.
                  \end{array}
               \right.  
\end{eqnarray*}
Since, the one stage cost and the density function 
$\phi_2({\bf y};\widetilde{\Pi}_{k-1})$ are time invariant, it is sufficient
to consider the class of stationary policies (see Proposition 2.2.2 of
\cite{books.bertsekas07b}). Let $\widetilde\mu:\mathcal{S}\to\mathcal{A}$ be a
stationary policy. Hence, the cost of using the policy $\widetilde\mu$ is given by 
\begin{eqnarray*}
J_{\widetilde\mu}(\pi_0) & = &  \EXP{\sum_{k=0}^\infty \widetilde c(\Pi_k,\widetilde\mu(\Pi_k)) \bigg\arrowvert \Pi_0 = \pi_0},
\end{eqnarray*}
and the optimal cost under the class of stationary policies is given by 
\begin{eqnarray*}
J^*(\pi_0) & = & \min_{\widetilde\mu} \EXP{\sum_{k=0}^\infty \widetilde c(\Pi_k,\widetilde\mu(\Pi_k))
\bigg\arrowvert \Pi_0 = \pi_0} 
\end{eqnarray*}
The DP that solves the above equation is given by the Bellman's
equation,
\begin{eqnarray*}
J^*(\pi) & = & \min \bigg\{\widetilde c(\pi,1), H_{J^*}(\pi) \bigg\}
\end{eqnarray*}
where
$H_{J^*}:[0,1]\to{\mathbb R}_+$ is defined as 
{\footnotesize
\begin{eqnarray*}
H_{J^*}(\pi) := \widetilde c(\pi,0) + \sum_{m=0}^n\gamma_m\EXPm{J^*\bigg(\Phi(\pi,(0,m),{\bf
Y})\bigg)}
\end{eqnarray*}
}
where ${\bf Y}$ and ${\bf y}$ are $m$--vectors.
The above equation can be written as 
\begin{eqnarray} 
\label{eqn:DP_for_control_q}
J^*(\pi)  
        & = & \min\left\{\lambda_f\cdot\big(1-\pi\big),  \pi + A_{J^*}\big({\pi}\big)\right\}.
\end{eqnarray}
where the function $A_{J^*}:[0,1]\to\mathbb{R}_+$ is defined as
\begin{align*} 
A_{J^*}({\pi}) 
= & \lambda_snq + \sum_{m=0}^n \gamma_m \EXPm{J^*\left(\frac{\tilde{\pi}\cdot\phi_1({\bf Y})}{\phi_2({\bf Y};\tilde{\pi})}\right)}. 
\end{align*}
The optimal policy $\mu^*$  that achieves $J^*$ gives the optimal
stopping rule, $\tau^*$. 
We now prove some properties of the optimal policy. 
\begin{theorem}
\label{thm:label5}
The optimal total cost function $J^*(\pi)$ is concave in $\pi$. 
\end{theorem}

\begin{theorem}
\label{thm:label6}
The optimal stopping rule is a threshold rule where the threshold is on
the a posteriori probability of change,
\begin{eqnarray*}
\tau^* & = & \inf\{k : \Pi_k \geqslant \Gamma\},
\end{eqnarray*}
for some $\Gamma \in [0, 1]$. The threshold $\Gamma$ depends on the
probability of false alarm constraint, $\alpha$ (among other parameters
like the distribution of $T$, $f_0$, $f_1$). 
\end{theorem}

For each $q \in [0,1]$, we compute the optimal mean detection delay
${\sf E_{DD}}$ (as a function of $q$), and then find the optimal $q^*$
for which the optimal mean detection delay is minimum.

\section{Numerical Results} 
\label{sec:numerical_results} 
\begin{figure}
\centering
\includegraphics[width=3.5in]{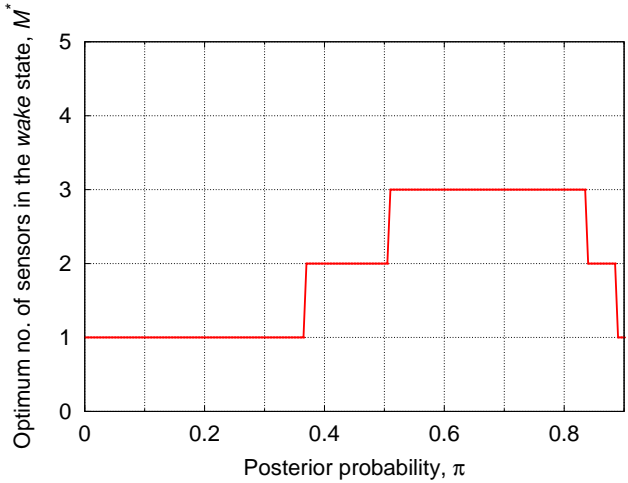}
\caption{Optimum number of sensors in the {\wake} state $M^*$ for $n = 10$ sensors, 
$\lambda_f = 100.0$, $\lambda_s = 0.5$, $f_0 \sim \mathcal{N}(0,1)$ and $f_1 \sim \mathcal{N}(1,1)$. Note that $\Gamma = 0.9$ corresponds to the threshold.}
\label{fig:Opt_control_M_M}
\end{figure}

We compute the optimal policy for each of the \sleep--${\sf wake}$
scheduling strategies given in Eqns. \ref{eqn:opt_M_staionary_policy},
\ref{eqn:DP_for_control_q_k}, \ref{eqn:DP_for_control_q} using
value--iteration technique (see \cite{books.bertsekas00a}). We consider
$n = 10$ sensors. The distributions of change--time $T$ is taken to be
geometric $(0.01)$ (and $\pi_0 = 0$). Also, the prechange and the
postchange distributions of the sensor observations are taken to be
$\mathcal{N}(0,1)$ and $\mathcal{N}(1,1)$. We set the cost per
observation per sensor, $\lambda_s$ to 0.5 and the cost of false alarm,
$\lambda_f$ to 100.0 (this corresponds to $\alpha =$ 0.04).

\begin{itemize}
\item {\bf Optimal control of $M_{k+1}$:}

We compute $M^*$ the optimal number of sensors to be in the
{\wake} state in time slot $k+1$ as a function of the a posteriori
probability of change $\pi$ (from the optimal policy $\mu^*$ given
by Eqn.\ref{eqn:opt_M_staionary_policy}) by the {\em value iteration}
algorithm 
\cite{books.bertsekas07b}, \cite{lerma} and plot in
Figure~\ref{fig:Opt_control_M_M}. We note that in any time slot, it is
not economical to use more than 3 sensors (though we have 10 sensors).
Also, from Figure~\ref{fig:Opt_control_M_M}, it is clear that $M^*$
increases monotonically for $\pi < 0.6$ and then decreases monotonically
for $\pi \ge 0.6$. Note that, the region $\pi \in [0.5, \ 0.82]$ 
requires many sensors for optimal detection whereas the region $[0.0, \
0.3]\cup[0.9, \ 1.0]$ requires the least number of sensors.  This is due
to the fact that {\em uncertainty} is more in the region $\pi \in [0.5,
\  0.82]$ whereas it is less in the region $[0.0, \ 0.3]\cup[0.9, \
1.0]$. 
   
\vspace{2mm}

In Figure~\ref{fig:simulation_control_M}, we plot the trajectory of a
sample path of $\Pi_k$ versus the time slot $k$. In our numerical
experiment, the event occurs at $T = 152$. When the number of sensors to
be in the {\wake} state $M_{k+1}$ is $M^*(\pi_k)$ (taken from 
Figure~\ref{fig:Opt_control_M_M}), for a threshold of 0.9, we see that 
the detection happens at $\tau_{M^*} = 161$. When $M_{k+1} = 10$ 
sensors (no {\sleep} scheduling), we find the detection epoch to be 
$\tau_{10} = 153$. When $M_{k+1} = 3$ sensors (we chose 3 because $M^* 
\le 3$), the stopping happens at $\tau_3 = 156$. From the above stopping
times, it is clear that the detection delay does not vary significantly
in the above three cases. By having an optimal {\sleep}--{$\sf wake$}
scheduling, we observe that until the event occurs only one sensor is in
{\wake} state and as soon as the event occurs, the {\sleep}--{$\sf wake$}
scheduler ramps up the number of sensors to 3, thereby making a quick
decision. Thus, the optimal {\sleep}--{$\sf wake$} scheduling uses a
minimal number of sensors before change and quickly ramps up the number
of sensors after change for quick detection.
Also, we see from
Figure~\ref{fig:simulation_control_M}, that the $\pi_k$ trajectory
corresponding to $M_{k+1}(\pi) =  10$ (and $M_{k+1}(\pi) =  3$) gives
more reliable information about the event than the $\pi_k$ trajectory
corresponding to $M_{k+1}(\pi) =  M^*$.    
\begin{figure}
\centering
\includegraphics[width=3.5in]{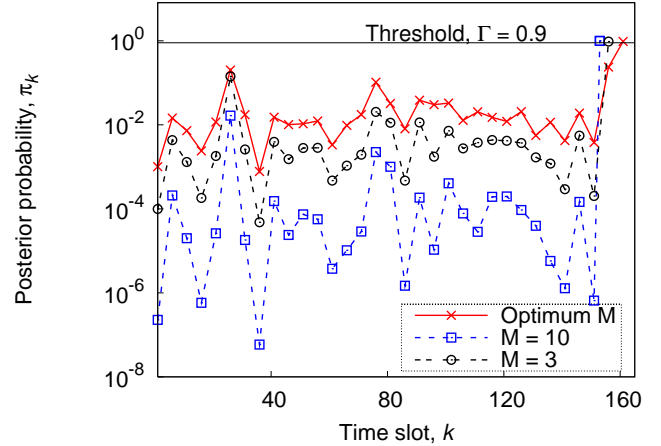}
\caption{A sample run of {\em event detection} with $n = 10$ sensors, 
$\lambda_f = 100.0$, $\lambda_s = 0.5$, $f_0 \sim \mathcal{N}(0,1)$ 
and $f_1 \sim \mathcal{N}(1,1)$.}
\label{fig:simulation_control_M}
\end{figure}
\begin{figure}
\centering
\includegraphics[width=3.5in]{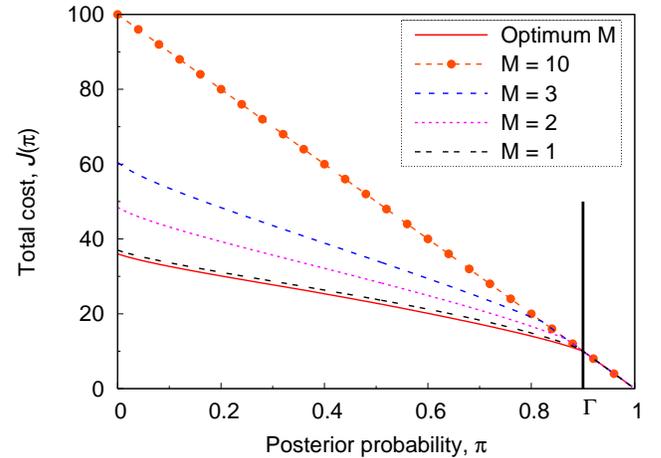}
\caption{Total cost $J(\pi)$ for $n = 10$ sensors, $\lambda_f = 100.0$, $\lambda_s = 0.5$, $f_0 \sim \mathcal{N}(0,1)$ and $f_1 \sim \mathcal{N}(1,1)$. Note that the threshold 
corresponding to $M = 1$ is 0.895, for $M = 2$ is 0.870, 
for $M = 3$ is 0.825, and for $M^*$ is $\Gamma = 0.9$.}
\label{fig:Opt_control_M_J}
\end{figure}

\begin{figure}[t]
\centering
\includegraphics[width=3.5in]{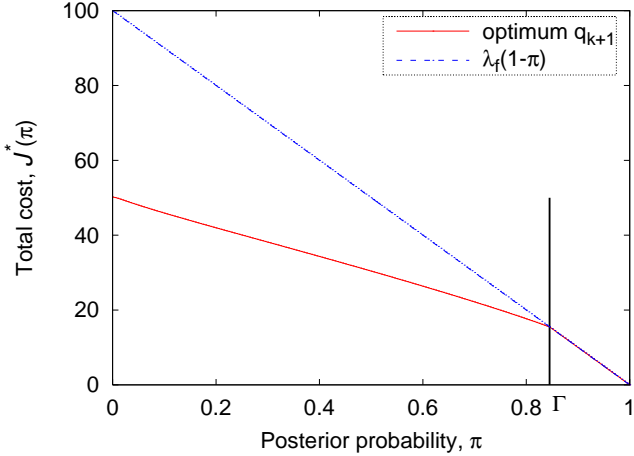}
\caption{Total cost $J^*(\pi)$ for $n = 10$ sensors, $\lambda_f = 100.0$,
$\lambda_s = 0.5$, $f_0 \sim \mathcal{N}(0,1)$ and $f_1 \sim
\mathcal{N}(1,1)$. The dashed line $\lambda_f(1-\pi)$ is the cost of
false alarm.}
\label{fig:Opt_control_q_J}
\end{figure}

We also plot the total cost function $J(\pi)$ for the above cases in 
Figure~\ref{fig:Opt_control_M_J}. Though the detection delays do not vary much, the total 
cost varies significantly. This is because the event happens at time slot 
$T = 152$. In the case of $M_{k+1} =  M^*$, it is clear from 
Figures~\ref{fig:Opt_control_M_M} and \ref{fig:simulation_control_M} that only one sensor 
is used for the first 158 time slots. This reduces the cost by 10 times compared to the
case of $M_{k+1} = 10$  (in this sample path) and about 3 times compared to the 
case of $M_{k+1} = 3$ (in this sample path). We note from Figure~\ref{fig:Opt_control_M_J}, 
that it is better to keep 3 sensors active all the time than keeping 10 
sensors active all the time. Also, in the case of $M_{k+1} = 1$, after
the event occurs, the a
posteriori probability takes more time to cross the threshold compared
to the optimal {\sleep}--{$\sf wake$} (which quickly ramps up from 1 to 3
sensors) and hence, the total cost corresponding to $M_{k+1}=1$ is slightly
worse than that of $M_{k+1} = M^*$.

\item {\bf Optimal control of $q_{k+1}$:}
In the case of control on $q_{k}$, we consider the same set of
parameters as in the case of control on $M_{k}$. We computed the optimal
policy from the DP defined in Eqn.~\ref{eqn:DP_for_control_q_k} by value
iteration. The optimal policy also gives the optimal probability of
choosing a sensor in the {\wake} state, $q^*_{k+1}$.  We plot the total
cost $J^*(\pi)$ in Figure~\ref{fig:Opt_control_q_J}. We also plot the
optimum probability of a sensor in the {\wake} state, $q^*(\pi)$ in
Figure~\ref{fig:Opt_control_q_q}. We observe that for $\pi \leq 0.72$,
$q^*(\pi)$ is an increasing function of $\pi$, and for $\pi > 0.72$,
$q^*(\pi)$ decreases with $\pi$. This agrees well with the intuition for
the optimal control on $M_{k+1}$.

\begin{figure}
\centering
\includegraphics[width=3.5in]{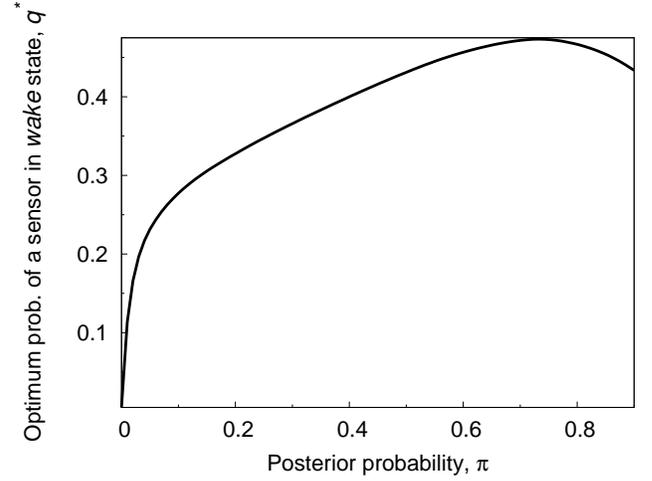}
\caption{Optimum probability of a sensor in the {\wake} state, $q_{k+1}^*(\pi)$ 
for $n = 10$ sensors, $\lambda_f = 100.0$, $\lambda_s = 0.5$,
$f_0 \sim \mathcal{N}(0,1)$ and $f_1 \sim \mathcal{N}(1,1)$.}
\label{fig:Opt_control_q_q}
\end{figure}

\begin{figure}
\centering
\includegraphics[width=3.5in]{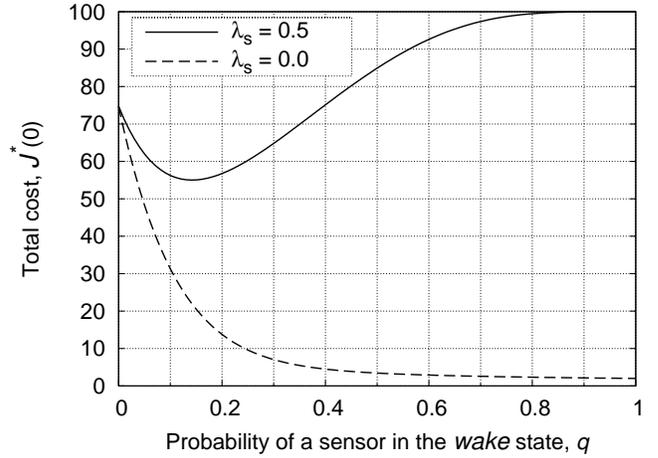}
\caption{Total cost $J^*(0)$ for $n = 10$ sensors, $\lambda_f = 100.0$,
$f_0 \sim \mathcal{N}(0,1)$ and $f_1 \sim \mathcal{N}(1,1)$.}
\label{fig:minimize_bayes_cost}
\end{figure}

\item {\bf Open loop control on $q$:}

We consider the same set of parameters for the case of open loop control
on $q$. We obtain $J^*(0)$ for various values of $q$ and plotted in the
Figure~\ref{fig:minimize_bayes_cost}.  We obtain the plot for $\lambda_s
= 0.5$ and for $\lambda_s = 0.0$. In the special case of $q=1$, i.e.,
having $M_{k+1} = 10$ sensors, and with $\lambda_s = 0.5$, we observe
that the total cost is 100 which matches with the corresponding cost in
Figure~\ref{fig:Opt_control_M_J}. Also, in the limiting case of $q \to
0$, all the sensor nodes are in the {\sleep} state at all time slots,
and the detection happens only based on Bayesian update (i.e., based on 
the prior distribution of $T$). Thus at $q = 0$, the total cost is the
same (which is 73) for $\lambda_s = 0.5$ and 
$\lambda_s = 0.0$ which is also evident from 
Figure~\ref{fig:minimize_bayes_cost}.

Note that when $\lambda_s > 0$, for
low values of $q$, the detection delay cost dominates over the
observation costs in $J^*(0)$ and for high values of $q$, the
observation costs dominate over the detection delay cost. Thus, there is
a trade--off between the detection delay cost and the observation costs
as $q$ varies. This is captured in the
Figure~\ref{fig:minimize_bayes_cost}. Note that the Bayesian cost is
optimal at $q = 0.15$.  When $\lambda_s = 0$, as $q$ increases the
detection delay decreases. Hence, we see the monotonically decreasing
trend for  $\lambda_s = 0.0$.       
\end{itemize}

From Figures~\ref{fig:Opt_control_M_J},~\ref{fig:Opt_control_q_J},
and \ref{fig:minimize_bayes_cost}, we note that the total cost
$J(\pi)$ is the least for optimal control on $M_{k+1}$. Also, we note
that in the open loop control case, the least total cost $J^*(0) = 55$
is achieved when the attempt probability, $q$ is $0.15$ 
(see Figure~\ref{fig:minimize_bayes_cost}; this
corresponds to an average of 1.5 sensors being {\sf active}). It is to be noted that
this cost is larger than that achieved by the optimal closed loop
policies ($J^*(0) = 50$ for the closed loop control on $q_{k+1}$ and
$J^*(0) = 38$ for the closed loop control on $M_{k+1}$). From
Figures~\ref{fig:simulation_control_M} and
\ref{fig:Opt_control_M_M}, we see that when $M_{k+1}(\pi) =
M^*(\pi)$, the switching of the sensors between {\sleep} and 
{\wake} states happen only in 2 slots out of 161 slots. Otherwise
only 1 sensor is on.

\section{Conclusion}
\label{sec:conclusion}
In this paper, we formulated the problem of jointly optimal
\sleep--${\sf wake}$ scheduling and event detection in a sensor network
that minimises the detection delay and the usage of
sensing/communication resources. We have set out to solve the problem in
Eqn.~\ref{eqn:bridge}. We have derived the optimal control for three
approaches using the theory of MDP. We showed the existence of the
optimal policy and obtained some structural results.  

We prescribe the \sleep--${\sf wake}$ scheduling policies as follows:
When there is a feedback between the fusion centre and the sensors and
if the feedback is unicast, it is optimal to use the control on
$M_{k+1}$ policy; when the feedback is only broadcast, then it is 
optimal to use the control on $q_{k+1}$. If there is no feedback between 
the fusion centre and the sensors, we prescribe the open loop control on 
$q$ policy.


\appendices
\appendix{\bf Proof of Theorem~\ref{thm:opt_m_J*_concave}}
\label{sec:appendix}

We use the following Lemma to prove
Theorem~\ref{thm:opt_m_J*_concave}.
\begin{lemma}
\label{lem:Lemma01}
If $f:[0,1] \to \mathbb{R}$ is concave, then for any ${\bf x} \in
{\mathbb R}^m$ (for any $m \in \mathbb{Z}_+$), the function 
$h:[0,1] \to \mathbb{R}$ defined  by 
\begin{align*} 
h(y) & = {\mathsf E}_{
\phi_2({\bf x};y)	
	}\left[f\left(\frac{y \phi_1({\bf X})}{y \phi_1({\bf X}) + (1-y)\phi_0({\bf X}) }\right) \right]   
\end{align*}
is concave in $y$, where $\phi_1({\bf x})$ and $\phi_0({\bf x})$ are pdfs on ${\bf X}$, and $\phi_2({\bf x};y) = y\phi_1({\bf x})+(1-y)\phi_0({\bf x})$.
\end{lemma}
\noindent
{\bf Proof } 
For any given ${\bf x}$, define the function $h_1:[0,1] \to \mathbb{R}$ as
\begin{eqnarray*}
&    & h_1(y; {\bf x})\\ 
& := & f\left(\frac{y \phi_1({\bf x})}{y \phi_1({\bf x}) + (1-y)\phi_0({\bf x}) }\right) \Big[y \phi_1({\bf x}) + (1-y)\phi_0({\bf x})\Big]. 
\end{eqnarray*}
As ${\mathsf T} := \int \cdots \ d{\bf x}$ is a linear operator and
$h(y) = \mathsf T h_1(y;{\bf x})$, it is sufficient to show that 
$h_1(y; {\bf x})$ is concave in $y$.
If $f(y)$ is concave then (see \cite{rockafellar})
\begin{align*} 
f(y) & = \inf_{(a_i, b_i) \in I} \big\{a_i y + b_i\big\}
\end{align*}
where $I = \{(a,b)\in\mathbb{R}^2: ay + b \ge f(y), y \in [0,1] \}$.  
Hence,
{
\begin{align*} 
& h_1(y;{\bf x}) \\
 = &f\left(\frac{y \phi_1({\bf x})}{y \phi_1({\bf x}) + (1-y)\phi_0({\bf x}) }\right)\Big[y \phi_1({\bf x}) + (1-y)\phi_0({\bf x})\Big] \\
 = & \inf_{(a_i, b_i) \in I} \left\{a_i \left(\frac{y \phi_1({\bf x})}{y
 \phi_1({\bf x}) + (1-y)\phi_0({\bf x}) }\right) + b_i\right\}\\
  & \cdot \Big[y \phi_1({\bf x}) + (1-y)\phi_0({\bf x}
)\Big] \\
 = & \inf_{(a_i, b_i) \in I} \left\{a_i y \phi_1({\bf x}) + b_i \Big[y \phi_1({\bf x}) + (1-y)\phi_0({\bf x})\Big] \right\} \\
 = & \inf_{(a_i, b_i) \in I} \left\{\Big((a_i+b_i) \phi_1({\bf x}) - b_i \phi_0({\bf x}
)\Big)y + b_i\phi_0({\bf x}) \right\} 
\end{align*}
}
which is an infimum of a collection of affine functions of $y$. This
implies that $h_1(y;{\bf x})$ is concave in $y$ (see
\cite{rockafellar}). 
$\hfill\blacksquare$

The optimal total cost function $J^*(\pi)$ can be computed using a {\em
value iteration} algortithm. Here, we first consider a finite
$K$--horizon problem and then we let $k \to \infty$, to obtain the
infinite horizon problem. 

Note that the cost--to--go function, $J_K^K(\pi) = 
\lambda_f\cdot\big(1 - \pi\big)$ is concave in $\pi$. Hence, by 
Lemma~\ref{lem:Lemma01}, we see that the cost--to--go functions  
$J_{K-1}^K(\pi),$ $J_{K-2}^K(\pi),$ $\cdots, J_0^K(\pi)$ are concave in
$\pi$. Hence for $0 \le \lambda \le 1$,
\begin{align*} 
J^*(\pi) & = \lim_{K \rightarrow \infty} J^K_0(\pi)\\
J^*(\lambda \pi_1 + (1-\lambda)\pi_2) & = \lim_{K \rightarrow \infty} J^K_0\Big(\lambda\pi_1+(1-\lambda)\pi_2\Big)\\
& \ge \lim_{K \rightarrow \infty} \lambda J^K_0(\pi_1)+ \lim_{K \rightarrow \infty} (1-\lambda)J^K_0(\pi_2)\\
& = \lambda J^*(\pi_1)+(1-\lambda)J^*(\pi_2)
\end{align*}
It follows that $J^*(\pi)$ is concave in $\pi$.
$\hfill\blacksquare$

\appendix{\bf Proof of Theorem~\ref{thm:policy-M}}

Define the maps $C:[0,1] \to \mathbb{R}_+$ and $H:[0,1] \to \mathbb{R}_+$, as 
\begin{align*} 
C(\pi)     & := \lambda_f\cdot\big(1 - \pi\big)\\
H(\pi)     & := \pi + A_{J^*}(\pi) 
\end{align*}
Note that $C(1) =  0$, $H(1) =  1$, $C(0) =   \lambda_f$ and $H(0) =   A_{J^*}(0)$.
Note that
\begin{eqnarray*} 
& &  A_{J^*}(0)\\
 & = & \min_{0 \le m \le n } \left\{\lambda_s m +  {\mathsf E}_{\SPHI{m}}\left[J^*\left(\frac{{p}\cdot\phi_1({\bf X}^{(m)})}{\phi_2({\bf X}^{(m)};p)}\right)\right] \right\}\\
 & \le & \min_{0 \le m \le n } \left\{\lambda_s m +  J^*\left({\mathsf
 E}_{\SPHI{m}}\left[\frac{p\cdot\phi_1({\bf X}^{(m)})}{\phi_2({\bf
 X}^{(m)};p)}\right]\right) \right\} \\
 &    = & \min_{0 \le m \le n } \left\{\lambda_s m + J^*\left(p\right) \right\}\\
 &    = & J^*\left(p\right)\\
&	 \le & \lambda_f\cdot\big(1 - p\big) \ \ \ \text{(from Eqn.~16)} 
\end{eqnarray*}
The inequality in the second step is justified using Jensen's inequality and 
the inequality in the last step follows from the definition of $J^*$.

Note that  $H(1) - C(1) > 0$ and $H(0) - C(0) <  0$. 
As the function $H(\pi) - C(\pi)$ is concave, by the {\em intermediate 
value theorem}, there exists $\Gamma \in [0,1]$ such that $H(\Gamma) = C(\Gamma)$.
This $\Gamma$ is unique as $H(\pi) = C(\pi)$ for at most two values of $\pi$. 
If in the interval $[0,1]$, there are two distinct values of $\pi$ for 
which $H(\pi) = C(\pi)$, then the signs of $H(0) - C(0)$ and  $H(1)-C(1)$
should be the same.
Hence, the optimal stopping rule is given by
\begin{align*} 
\tau^* & = \inf\left\{k: \Pi_k \ge \Gamma\right\}
\end{align*}
where the threshold $\Gamma$ is given by $\Gamma + A_{J^*}(\Gamma) = \lambda_f\cdot\big(1 - \Gamma\big)$.

$\hfill\blacksquare$

\appendix{\bf Proof of Theorem~\ref{thm:monotone}}

Define 
\begin{eqnarray*}
\phi_j({\bf x}^{(m)}) & := &\prod_{i=1}^m f_j(x^{(i)}), \ j = 0,1.\\ 
{\bf x}^{(l)}         & := &(x^{(1)},   x^{(2)}, \cdots, x^{(m)}, x^{(m+1)}, \cdots, x^{(l)})\\
{\bf u}               & := &(x^{(1)},   x^{(2)}, \cdots, x^{(m)})\\
{\bf v}               & := &(x^{(m+1)}, x^{(m+2)}, \cdots, x^{(l)})\\
\hat{\pi} & := &\frac{\tilde{\pi} \phi_1({\bf u})}{\tilde{\pi} \phi_1({\bf u}) + (1-\tilde{\pi}) \phi_0({\bf u})}
\end{eqnarray*}
Note that
\begin{eqnarray*}
& & B_{J^*}^{(l)}({\pi}) \\
& = &  \int_{{\mathbb R^l}}J^*\left( \frac{{\tilde{\pi}}\cdot\phi_1({\bf x}^{(l)})}{\phi_2({\bf x}^{(l)};\tilde{\pi})}\right)\Big[ \PPHII{l} \Big] \ d{\bf x}^{(l)}\\ 
 & = & \int_{{\mathbb R^m}} \int_{\mathbb R^{l-m}} J^*\left(\frac{\hat{\pi} \phi_1({\bf v})} {\BALA{v}{\hat{\pi}}} \right)\BALA{v}{\hat{\pi}} \ d{\bf v}  \BALA{u}{\tilde{\pi}} \  d{\bf u} \\
 & \le & \int_{{\mathbb R^m}} J^*\left(\int_{\mathbb R^{l-m}} \frac{\hat{\pi} \phi_1({\bf v})} {\BALA{v}{\hat{\pi}}} \Big[\BALA{v}{\hat{\pi}}\Big]  d{\bf v} \right) \BALA{u}{\tilde{\pi}} \  d{\bf u} \\
  &= & \int_{{\mathbb R^m}} J^*\left(\hat{\pi}\right) \BALA{u}{\tilde{\pi}}  d{\bf u} \\
& = & B_{J^*}^{(m)}({\pi})
\end{eqnarray*}
As $J^*$ is concave, the inequality in the second line follows from Jensen's
inequality. Hence proved.
$\hfill\blacksquare$

\appendix{\bf Proof of Theorem~\ref{thm:label1}}

Eqn.~\ref{eqn:OPT_CONTROL_M_A_J_STAR} and the monotone property of $d(m;.)$ 
proves the theorem.  
$\hfill\blacksquare$

%
%

\appendix{\bf Proof of Theorem~\ref{thm:label3}}

Follows from the proof of Theorem~\ref{thm:opt_m_J*_concave}.
$\hfill\blacksquare$

\appendix{\bf Proof of Theorem~\ref{thm:policy}}

Follows from the proof of Theorem~\ref{thm:policy-M}.
$\hfill\blacksquare$

%
%

\appendix{\bf Proof of Theorem~\ref{thm:label5}}

Follows from the proof of Theorem~\ref{thm:opt_m_J*_concave}.
$\hfill\blacksquare$

\appendix{\bf Proof of Theorem~\ref{thm:label6}}

Follows from the proof of Theorem~\ref{thm:policy-M}.
$\hfill\blacksquare$
\ifCLASSOPTIONcaptionsoff
  \newpage
\fi



\bibliographystyle{IEEEtran}
\bibliography{IEEEabrv,premkumar_kumar11sleep_wake_scheduling}  

%








\end{document}